\documentclass[iop]{emulateapj}

\def\ergsec{\hbox{erg s$^{-1}$}} 
\def\ergcm{\hbox{erg s$^{-1}$ cm$^{-2}$}}

\def\xmm{\emph{XMM-Newton}}
\def\chandra{\emph{Chandra}}

\usepackage{multirow}

\shorttitle{ChaMPlane Survey: Identification of a new cataclysmic variable}
\shortauthors{Servillat et al.}

\begin{document}


\title{Spectroscopic Follow-up of X-ray Sources in the ChaMPlane Survey: \\Identification of a New Cataclysmic Variable}


\author{M. Servillat\altaffilmark{1}, J. Grindlay\altaffilmark{1}, M. van den Berg\altaffilmark{2,1}, J. Hong\altaffilmark{1}, P. Zhao\altaffilmark{1}, B. Allen\altaffilmark{1}}
\affil{\altaffilmark{1}Harvard-Smithsonian Center for Astrophysics, 60 Garden Street, MS-67, Cambridge, MA 02138
\email{mservillat@cfa.harvard.edu}}
\affil{\altaffilmark{2}Astronomical Institute, Utrecht University, Princetonplein 5, 3508 TA, The Netherlands}




\begin{abstract}
{We present a multi-object optical spectroscopy follow-up study of X-ray sources in a field along the Galactic Plane ($l=327.42^{\circ}$, $b=2.26^{\circ}$) which is part of the \chandra\ Multi-wavelength Plane survey (ChaMPlane). We obtained spectra for 46 stars, including 15 likely counterparts to X-ray sources, and sources showing an H$\alpha$ color excess. 
This has led to the identification of a new cataclysmic variable (CV), CXOPS J154305.5-522709, also named ChaMPlane Bright Source 7 (CBS~7), and we identified 8 X-ray sources in the field as active late-type stars.
CBS~7 was previously studied in X-rays and showed a hard spectrum and two periods: $1.22\pm0.08$~h and $2.43\pm0.26$~h. We present here clear evidence that the source is a CV through the detection of H, HeI and HeII emission lines in its optical spectrum. 
The hard X-ray spectrum and the presence of the HeII $\lambda$4686 in emission with a large equivalent width suggest a magnetic CV. 
The near-infrared counterpart is significantly variable, and we found a period consistent with the longest X-ray period at $2.39\pm0.05$~h but not the shortest X-ray period.
If this period is the orbital period, this would place the system in the CV period gap. 
The possible orbital period suggests a dM4$\pm$1 companion star. 
The distance is then estimated to be $\sim$1~kpc. 
The system could be a relatively hard and X-ray luminous polar or an intermediate polar, possibly nearly-synchronous. }
\end{abstract}






\section{Introduction}



Cataclysmic variables (CVs) are semi-detached binary stars with orbital periods typically of the order of hours, consisting of a white dwarf (WD) primary accreting via Roche lobe overflow from a companion star which is usually a late-type, main-sequence star. Subtypes of CVs include non-magnetic CVs where an accretion disk forms around the primary, and magnetic systems (MCVs) where the white dwarf magnetic field is strong enough to truncate (intermediate polars --- IPs --- with $10^5<B<10^7$~G) or disrupt (polars, $B>10^7$~G) the accretion disk. The mass flow is then channeled along the magnetic field lines and creates a shock at the magnetic pole of the WD \citep[e.g.][]{Cropper:1990p816,Patterson:1994p1017}.
{Polars are identified through their polarized optical emission, due to cyclotron emission processes near the surface of the WD.}
Polars are synchronous, locked systems, but the magnetic field in IPs is generally not strong enough to tidally lock the WD to the companion star and IPs are asynchronous rotators.
IPs are thus often characterized by two distinct periods in their X-ray lightcurves. The first is associated with the binary orbital period ($P_{orb}$), and the second, shorter, with the spin period of the WD ($P_{spin} \ll P_{orb}$). IPs tend to have larger orbital periods than polars, and the latter could be the end-products of the evolution of part of the IPs \citep[e.g.][]{Norton:2008p4760}.

The sample of known field CVs suffers from various selection effects \citep[e.g.][]{Pretorius:2007p4659,Servillat:2011p4345}, and is likely biased towards bright and highly variable systems. 
{Population synthesis models indicate that a fainter population showing fewer outbursts may dominate the population of DNe (up to 70\%, \citealt{Kolb:1993p1211}; \citealt{Howell:1997p1209}). Recent detections in the SDSS of faint CVs with short periods confirmed the existence of such a dominant population \citep{Gansicke:2009p1204}, but its proportion in absolute numbers is not fully constrained. The overall Galactic CV population, and the demographics of the various subclasses of CVs, are thus not fully understood, and this is especially true in the obscured regions of the Galactic Plane.}

In order to obtain an unbiased sample, we conduct the \chandra\ Multi-wavelength Plane Survey (ChaMPlane) which allows the identification of serendipitous X-ray sources discovered by the \chandra\ X-ray Observatory along the Galactic Plane \citep{Grindlay:2005p738}.
Based on a multi-wavelength dataset, we aim to identify rare Galactic populations of objects such as accreting white dwarfs, neutron stars, and black holes in order to study the nature, distribution and evolution of accreting compact objects in the Galaxy.

ChaMPlane has focused on the dense population of thousands of unidentified X-ray sources detected in the Galactic Bulge \citep[e.g.][]{Muno:2009p875,Hong:2009p2767} with deep observations of three low extinction windows (Baade's window at latitude $b=-3.8^{\circ}$, Stanek's window at $-2.2^{\circ}$ and Limiting Window at $-1.4^{\circ}$, South of the Galactic Center). This led to the discovery of a significant population of candidate CVs \citep{vandenBerg:2006p5431,VanDenBerg:2009p1014,Hong:2009p2775,Koenig:2008p5597}, suggesting the presence of a large number of MCVs in the Galactic Bulge.
We recently identified 10 periodic sources in the Limiting Window \citep{Hong:2011p4354}. These sources seem consistent with IPs based on their X-ray luminosity and spectra, but they are more likely to be polars based on their detected periods (1 to 3~h). 
{These puzzling properties could be explained if the sources are polars with hard X-ray spectra, or members of a rare sub-class of MCVs, nearly synchronous IPs ($P_{spin}/P_{orb}>0.3$, 6 known sources, \citealt{Ritter:2003p2684,Ritter:2010p2694}), some of which are considered as the missing link in the evolution of the MCVs from IPs to polars.}

We selected a subset of the brightest unidentified X-ray sources in the ChaMPlane database in order to characterize the Galactic Plane population of X-ray sources (\citealt{vandenBerg:2011p7449}; \citealt{Penner:2008p376}). One of these sources, CXOPS J154305.5-522709 or ChaMPlane Bright Source 7 (CBS~7) was tentatively classified as a CV based on its X-ray properties and H$\alpha$ color excess.
CBS~7 was found in the field of the \chandra\ archived observation of the high mass X-ray binary pulsar 4U 1538--52 (ObsId 90, $l=327.42^{\circ}$, $b=2.26^{\circ}$) which is part of ChaMPlane.

We report here new multi-object optical spectroscopy observations of selected X-ray sources in this field and near-infrared (NIR) images centered on CBS~7 in Section~\ref{data}. We present results for CBS~7 confirming the source as a CV in Section~\ref{results} and discuss in more detail its nature in Section~\ref{discuss}.

\section{Prior observations of CBS~7}
\label{prior}

The X-ray source CBS~7 is described by van den Berg et al. (2011).
It was observed for 24~ks on 2000 April 8 at UT 22:45:13 with the ACIS-I instrument onboard \chandra\ (ObsID 90). 
A total of 65 sources were detected with the ChaMPlane processing pipeline \citep{Hong:2005p6560} in this field.
CBS~7 was found to be periodic with two possible periods detected in the X-ray dataset: $P_{1} \sim 4392\pm290$~s ($1.22\pm0.08$~h) and $P_{2} \sim 8772\pm957$~s ($2.43\pm0.26$~h).
The 0.3--8 keV unabsorbed flux is $5.1\pm0.2\times10^{-13}$~\ergcm.
{The source was also observed with \xmm\ on 2003 Aug 14 (ObsID 0152780201; 81 ks).}
The X-ray spectrum of CBS~7 (2XMM J154305.5-522709, \citealt{Watson:2009p1813}) is well modeled by an absorbed power law with an absorption $N_{\mathrm{H}}=2.4\pm0.2\times10^{21}$~cm$^{-2}$ and a photon index $\Gamma=1.30\pm0.06$, and possibly the addition of an Fe line (96\% confidence level). 
{Using a bremsstrahlung model, the temperature of the plasma is constrained to be $>$30~keV.}
The distance derived from the $N_{\mathrm{H}}$ value, assuming it comes primarily from material along the line of sight, gives a {distance estimate of $1.4\pm0.1$~kpc and a luminosity of $\sim10^{32}$~\ergsec\ (van den Berg et al. 2011)}.


\begin{figure}
\centering
\includegraphics[width=\columnwidth]{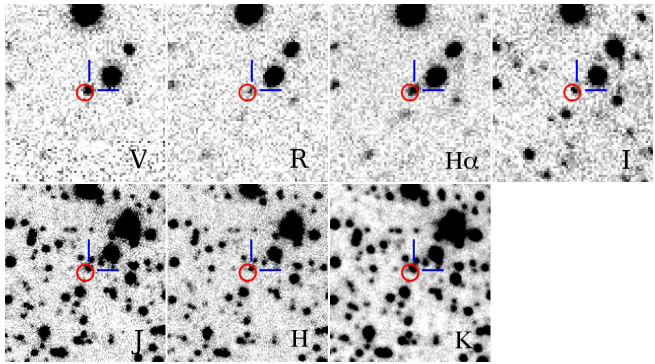}
\caption{Thumbnails images of optical observations (top row) and near-infrared observations (bottom row) of CBS~7. The \chandra\ X-ray error circle of CBS~7 (0.7\arcsec\ at 95\%) is shown on each image, and the likely counterpart of the X-ray source is indicated by lines. Images are 20'' large with North up and East left.}
\label{fig:im}
\end{figure}

\begin{deluxetable}{ccccc}
\tablecaption{Log of optical observations of CBS~7.\label{tab:optmag}}
\tablewidth{0pt}
\tablehead{\colhead{Date (UT)} & \colhead{Filter} & \colhead{Exp.Time} & \colhead{Airmass}}
\startdata
2001-05-18T04:16:43 & $V$          & 180s  & 1.08   \\
2001-05-18T01:42:41 & $R$          & 240s  & 1.29   \\
2001-05-18T04:26:52 & $I$           & 180s  & 1.08   \\
2001-05-18T02:23:20 & H$\alpha$ & 1500s & 1.20  
\enddata 
\end{deluxetable}

The ChaMPlane X-ray Optical Plane Survey (CXOPS) now covers 74 \chandra\ fields and in particular the field containing 4U 1538--52 \citep[field 16 of][]{Zhao:2005p4431} where CBS~7 was found. This field was observed on 2001 May 18 with the CTIO Blanco 4m Telescope and the Mosaic instrument using the $V$, $R$, H$\alpha$ and $I$ bands (see Table~\ref{tab:optmag}). The source in the \chandra\ error circle of CBS~7 shows an excess in the H$\alpha$ narrow band filter (H$\alpha$--$R=-0.5\pm0.2$) at R.A. = 15$^{h}$43$^{m}$05.51$^{s}$ and Dec = --52$^{\circ}$27\arcmin09.6\arcsec. This counterpart is shown in Figure~\ref{fig:im} (top row) for each band. 

We found 48 possible optical identifications after cross-matching the X-ray and optical source lists to find counterparts candidates within the 95\% error circle taking into account the errors on the optical/X-ray boresight. 
We repeated 24 times this process after applying shifts of 5\arcsec\ between the 2 source lists following a squared grid pattern. This showed that $27.5\pm2.1$ chance alignments are expected for the 65 X-ray sources, so that $\sim$20 of the 48 optical counterpart candidates may be real.

\section{Data reduction}
\label{data}

\subsection{Multi-object optical spectroscopy}

We observed the ChaMPlane field containing CBS~7 in multi-object slit spectroscopy mode with the Inamori-Magellan Areal Camera \& Spectrograph (IMACS, \citealt{Dressler:2011p4474}) at the Baade Magellan Telescope on 2010 Aug 30 with the f/4 camera and a 300 lines/mm grating (3650--9740~\AA, 0.743~\AA/pixel). We obtained 3 observations of 30 min (hereafter sp1, sp2 and sp3) for 46 targets. We also obtained arc-lamp exposures after each spectrum acquisition and bias, dark and flat fields at the beginning of the night.

The selection of targets was based on their X-ray or optical properties. Prime candidates are counterparts of X-ray sources with an H$\alpha$ excess (4 targets, including CBS~7), counterparts to any of the X-ray sources (11 targets), and stars with an H$\alpha$ excess but no X-ray emission (31 targets). We define here an H$\alpha$ excess as a H$\alpha$--$R$ color less than --0.2 in magnitude. This selection is similar to the selection performed by \citet{Koenig:2008p5597} for 5 fields in the Galactic Bulge.

We processed the data with the COSMOS v2.16 software\footnote{http://obs.carnegiescience.edu/Code/cosmos} and obtained roughly calibrated 2D spectra for each target. We then used the IRAF v2.14 \citep{Tody:1993p4441} onedspec and twodspec packages (apall function) to refine the wavelength calibration (giving a root mean square error of 0.2 \AA) and to perform the optimal extraction of spectra for each target. The full width half maximum (FWHM) of the arc-lamp emission lines and the sky lines is 5~\AA. The flux calibration was performed using the observations of the standard stars EG21 and LTT 7987 obtained the same night.

\subsection{Near-infrared photometry}

\begin{deluxetable}{ccccc}
\tablecaption{PANIC observation sequences and magnitudes for CBS~7.\label{tab:ispiobs}}
\tablewidth{0pt}
\tablehead{\colhead{Date (UT)} & \colhead{Filter} & \colhead{Sequence} & \colhead{Airmass} & \colhead{Mag}
}
\startdata
2010-06-30T23:09:30 & $J$   & 5x4x30s & 1.26 & 19.01$\pm$0.18  \\[0.5ex] \tableline \\[-2ex]
2010-06-30T23:25:55 & $H$  & 5x6x20s & 1.23 & 18.02$\pm$0.15  \\[0.5ex] \tableline \\[-2ex]
2010-06-30T22:58:37 & \multirow{2}{*}{$K_s$}  & \multirow{2}{*}{5x6x10s} &  \multirow{2}{*}{1.20--1.70} &  \multirow{2}{*}{16.77--17.94}  \\
2010-07-01T06:20:22 &  &  & 
\enddata
\tablecomments{A total of 7 $K_s$-band sequences were completed, hence we indicate the range of dates, airmass and magnitudes for these observations. Errors on the $K_{s}$-band magnitudes are $\pm$0.12, as can be seen on the light-curve in Figure~\ref{fig:lck}}
\end{deluxetable}

We obtained NIR images centered on CBS~7 with the Persson’s Auxiliary Nasmyth Infrared Camera (PANIC, \citealt{Martini:2004p3402}) on the Baade Magellan Telescope at Las Campanas, Chile. The observations were performed on 2010 June 30 and cover 8h. We used a 5-point dithering pattern in order to estimate and subtract the contribution from the sky in the images.

We processed the images with IRAF and developed pyRAF\footnote{http://www.stsci.edu/resources/software\_hardware/pyraf/} v1.9 procedures based on the PANIC package \citep{Martini:2004p3402}. The images were bias subtracted and flat-fielded. We then estimated the contribution of the sky using a two step procedure: (1) by computing the median of all dithered position images, (2) by masking any star remaining in the sky image, re-computing the median image and interpolating over the masked region. This method is particularly efficient for crowded fields where the median alone cannot remove all the emission from stars in the field. We then shifted all the images to align them on the first dithered position and stacked them.

We added an astrometry solution to the image using SCAMP \citep{Bertin:2006p3072}. We first used the approximate coordinates written in the file header during the acquisition and used a first degree polynomial to fit the detected stars to the 2MASS catalog of point sources \citep{Skrutskie:2006p4390}. This leads to a positional precision of $0\farcs2$ at 1$\sigma$. 

We performed the photometry extraction for each image with Sextractor \citep{Bertin:1996p4473}. We used the 2MASS catalog as a reference (Vega magnitudes) to estimate the zero point for each filter and each image. This leads to a median error on the magnitudes of 0.15.


\section{Results}
\label{results}

\subsection{Near-infrared counterpart to CBS~7}

\begin{figure}
\centering
\includegraphics[width=.8\columnwidth]{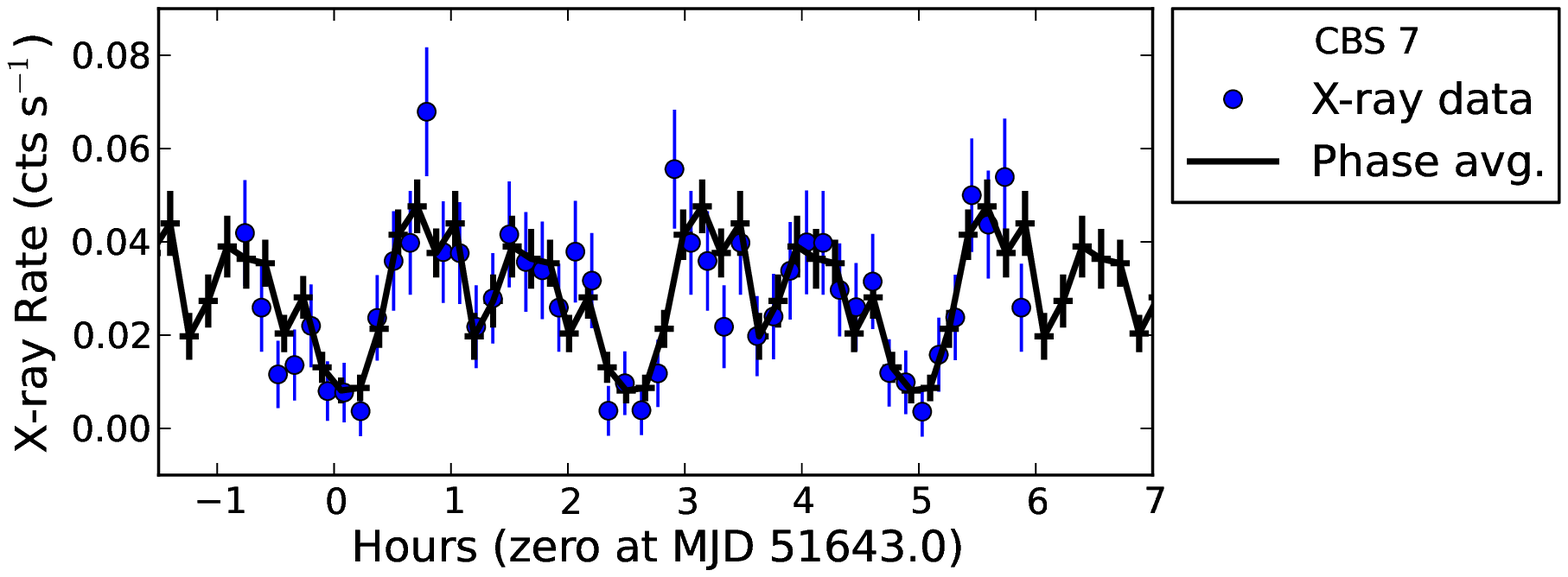}\\
\includegraphics[width=.8\columnwidth]{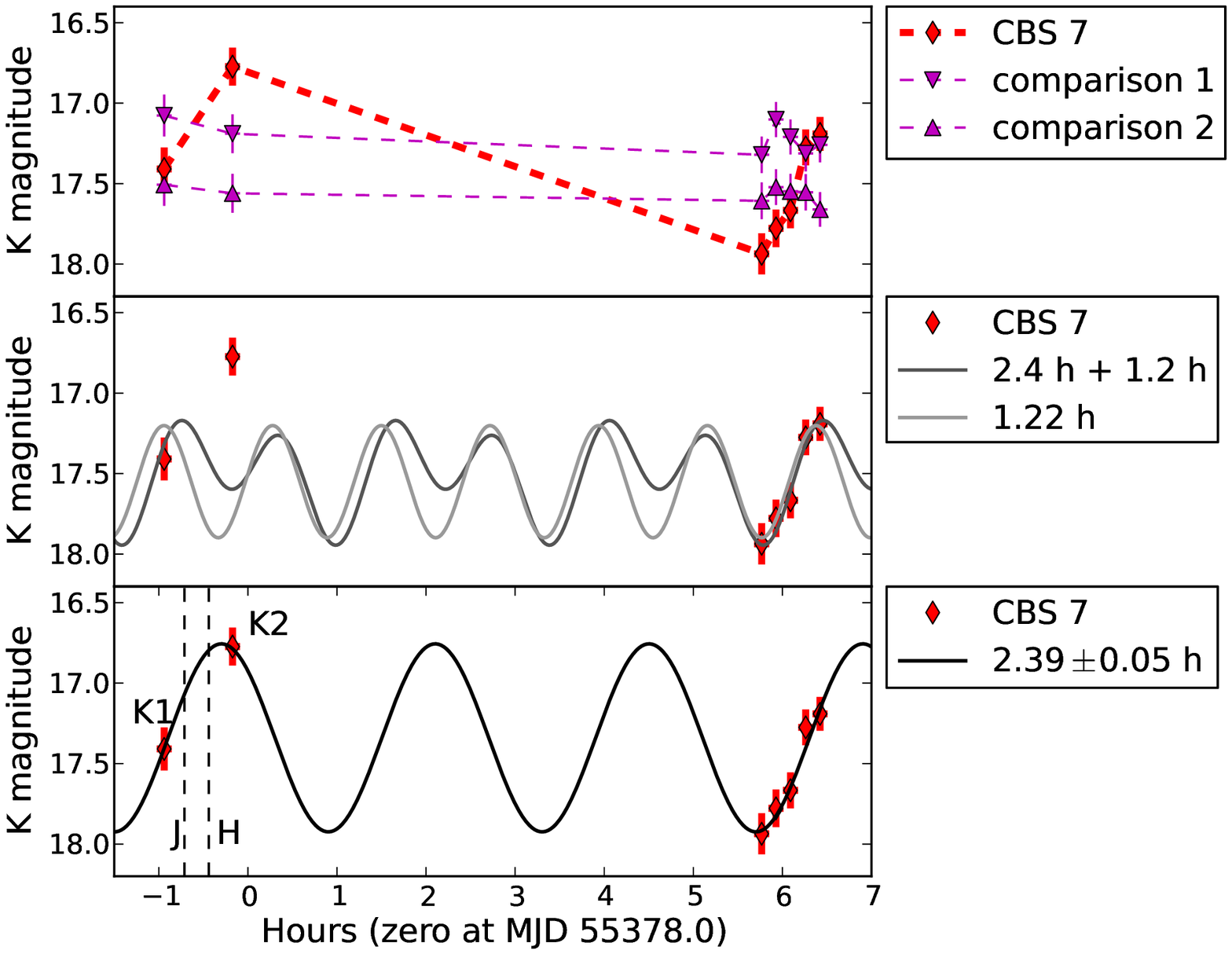}
\caption{Light-curves of CBS~7. Upper panel: X-ray lightcurve from \chandra, phase averaged (using the period $P_{2}=2.39$~h) and repeated over several periods to cover a similar range as the $K_s$-band observations. Lower panels: $K_s$-band lightcurve obtained with PANIC. Comparison with two reference stars (top). Fit with $P_{1}\sim1.2$~h or a combination of $P_{1}$ and $P_{2}\sim2.4$~h (middle). Fit with $P_{2}$ (bottom). {We note that other periods are consistent with the data (see text).} Vertical dashed lines indicate the mid exposure time of $J$ and $H$ band observations. Two $K_{S}$ observations (K1 and K2) are used in color determination (see Figure~\ref{fig:ccd}).}
\label{fig:lck}
\end{figure}

\begin{figure}
\centering
\includegraphics[width=\columnwidth]{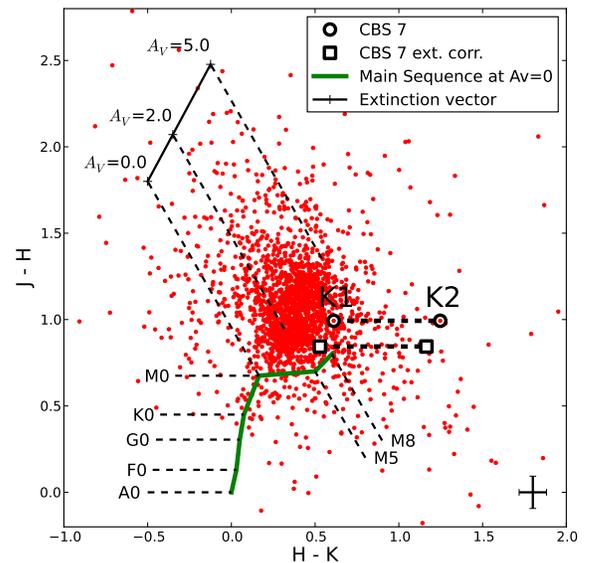}
\caption{Color-color diagram using NIR bands $J$, $H$ and $K_s$. The main sequence with no extinction is indicated as a thick green line \citep[from][]{Bessell:1988p4623}. A vector indicates the effect of extinction for a dM0 star and $A_{V}$=0, 2, 5 mag, corresponding to distances $d$~=~0, 1.9, 4.5~kpc (using \citealt{Drimmel:2003p4664} for conversions). The median error of the colors is shown at the bottom right of the plot. Two points are reported for CBS~7 in each case (direct value and extinction corrected values), using respectively the K1 and K2 measurements in the $K_s$-band (see Figure~\ref{fig:lck}).}
\label{fig:ccd}
\end{figure}

We looked for variability in the $K_s$-band using the 7 different images obtained. The observations were designed to cover {3 times the longest X-ray period of CBS~7 with a close sampling over half a period in order to look for variability in this band}. We found that the NIR counterpart aligned with the optical counterpart showed significant variability in the $K_s$-band. Figure~\ref{fig:lck} shows the target lightcurve compared with 2 nearby comparison stars with similar magnitude.

We fitted the lightcurve with a sinusoid of fixed period $P_{1}$ or a combination of $P_{1}$ and $P_{2}$ which mimics the X-ray lightcurve (Figure~\ref{fig:lck}). These periods can be ruled out in the $K_s$-band as they cannot explain the second $K_s$ measurement (K2 in Figure~\ref{fig:lck}). However, we obtained a convincing fit with the period $P_{2}$ only. We then let the period be free in the fit and constrained the period to $P_{2}=2.39\pm0.05$~h ($\chi^{2}=1.4$ for 3 degrees of freedom) with a mean Vega $K_s$ mag of $17.34\pm0.14$ and an amplitude of $0.58\pm0.12$ mag (Figure~\ref{fig:lck}).
{We further investigated the periods allowed by the K-band lightcurve, and found that sinusoids with periods of 1.79$\pm$0.03, 3.6$\pm$0.1 and 7.2$\pm$0.2~h can also fit the $K_s$-band lightcurve ($\chi^{2}$ of 4.4, 1.4 and 1.6, respectively, for 3 degrees of freedom). Shorter periods seem unlikely, mainly because they cannot reproduce the K2 measurement (${\chi^{2}>10}$). It is also possible that the variations observed are not periodic, but due to flares.}

We show in Figure~\ref{fig:ccd} the color-color diagram of the stars in the field of view using the $JHK_s$ bands. We plotted the main sequence with no extinction and a vector indicating the effect of extinction using R=3.1 \citep{Savage:1979p5573}. Using the \citet{Drimmel:2003p4664} model of  distribution of the dust in the Galaxy for the position of CBS 7, the extinction values $A_{V}$~=~0, 2, 5 mag convert to distances $d$~=~0, 1.9, 4.5~kpc, and the maximum extinction along the line of sight is $A_{V}=7.0$.
For CBS~7, the X-ray absorption $N_{\rm H}=2.4\pm0.2\times10^{21}$~cm$^{-2}$ converts \citep{Predehl:1995p5533} to an extinction along the line of sight of $A_{V}=1.3\pm0.2$, which we use to correct the colors in Figure~\ref{fig:ccd}.

Due to the variability of the target, there is a possible additional error on the color estimation. We thus computed 2 colors using the K1 and K2 measurements (as labelled in Figure~\ref{fig:lck}). The real value should lie between these two points. We note that the $J-H$ color could also be uncertain by $\sim0.2$~mag due to the small separation in time of the observations.
CBS~7 appears to be redder than the bulk of sources in the same field. After de-reddening, it could correspond to a M3 main sequence star or possibly less massive main sequence star.
{The $J$, $H$, K1 and K2 magnitudes were not obtained during the faintest phase, when the contribution from the companion to the combined emission is expected to be the largest. There are thus uncertainties in the estimation of colors for the secondary star that are difficult to assess with the data at hand (variability of the source, likely contribution of the accretion disk component, and reddening along the line of sight).}

\subsection{Optical spectra of CBS~7}

\noindent
\begin{deluxetable}{ccccccc}
\tablecaption{Emission line parameters for CBS~7.\label{tbl:el}}
\tablewidth{0pt}
\tablehead{\colhead{Line} & \colhead{Obs} & \colhead{$\lambda_{\mathrm{obs}}$} & \colhead{EW} & \colhead{FWHM} & \colhead{Flux}  & \colhead{$v_{\mathrm{radial}}$}  \\
 (1) & (2) & (3) & (4) &  (5) & (6) & (7)
}
\startdata
H$\alpha$ & sp1 & 6562.0 & -64.9 & 23.9 & 88.1 & -38.0 \\
          & sp2 & 6563.3 & -83.2 & 24.9 & 58.7 & 22.9 \\
          & sp3 & 6564.5 & -128.6 & 35.4 & 68.2 & 78.3 \\
[0.5ex] \tableline \\[-2ex]
H$\beta$ & sp1 & 4860.8 & -59.3 & 21.6 & 64.2 & -31.6 \\
         & sp2 & 4864.2 & -62.1 & 21.0 & 42.2 & 174.8 \\
         & sp3 & 4862.3 & -90.0 & 23.3 & 49.8 & 58.0 \\
[0.5ex] \tableline \\[-2ex]
H$\gamma$ & sp1 & 4340.0 & -58.0 & 25.6 & 42.1 & -33.2 \\
          & sp2 & 4344.7 & -52.9 & 25.8 & 23.8 & 288.7 \\
          & sp3 & 4342.1 & -61.7 & 22.2 & 22.2 & 109.7 \\
[0.5ex] \tableline \\[-2ex]
HeII $\lambda$4686 & sp1 & 4686.2 & -22.3 & 14.5 & 23.1 & 33.9 \\
                   & sp2 & 4690.0 & -27.6 & 15.7 & 18.1 & 273.2 \\
                   & sp3 & 4684.7 & -49.0 & 22.6 & 25.9 & -63.1 \\
[0.5ex] \tableline \\[-2ex]
HeI $\lambda$6678 & sp1 & 6676.5 & -9.9 & 14.5 & 13.5 & -65.3 \\
                  & sp2 & 6678.5 & -10.2 & 16.1 & 7.1 & 20.8 \\
                  & sp3 & 6677.9 & -12.2 & 18.8 & 6.4 & -6.1
\enddata 
\tablecomments{Columns: (1) Name of the emission line; (2) Label of spectrum observed; (3) Wavelength of the line in \AA; (4) Equivalent width in \AA; (5) Full width half maximum in \AA; (6) Flux of the line in $10^{-17}$~\ergcm~\AA$^{-1}$; (7) Radial velocity in km~s$^{-1}$.}
\end{deluxetable}

\begin{figure*}[tb]
\centering
\includegraphics[width=\textwidth]{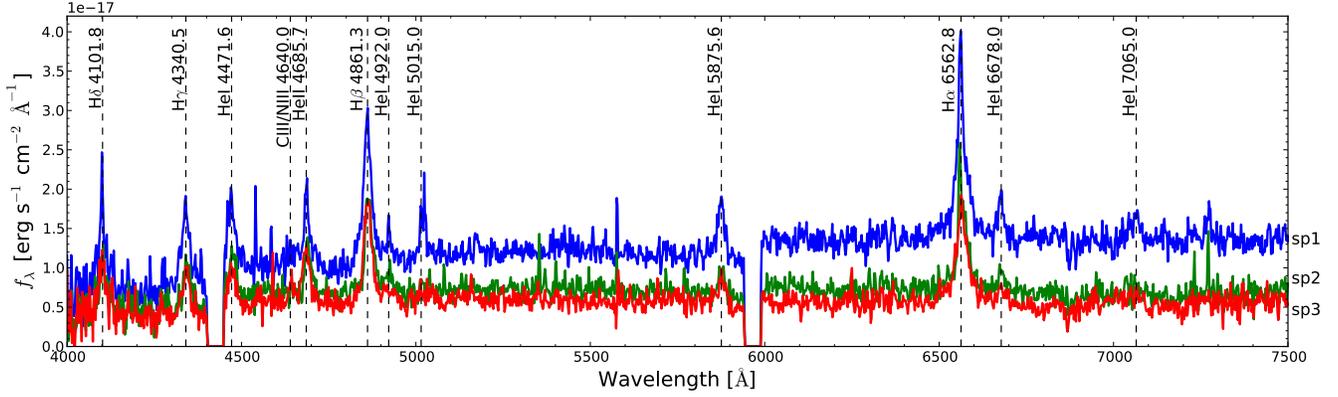}
\caption{Spectra of CBS~7 obtained with IMACS. The spectra are not de-reddened. Labels for each spectrum are indicated on the right vertical axis. Dashed lines indicate the position of H, HeI and HeII lines identified in the spectra.}
\label{fig:3sp}
\end{figure*}

\begin{figure}
\centering
\includegraphics[width=\columnwidth,angle=0]{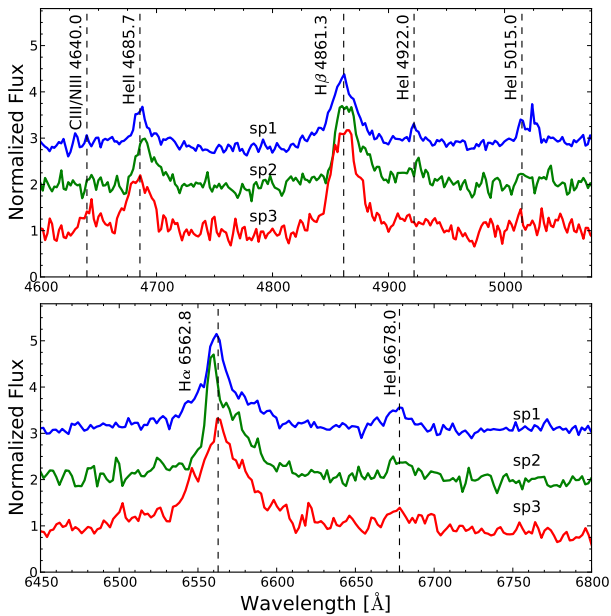}
\caption{Close-up on emission lines for CBS~7. The spectra continuum were normalized to 1 and an arbitrary offset of 1 has been applied between each spectrum for clarity.}
\label{fig:3spzoom}
\end{figure}

\begin{figure}[h]
\centering
\includegraphics[width=\columnwidth]{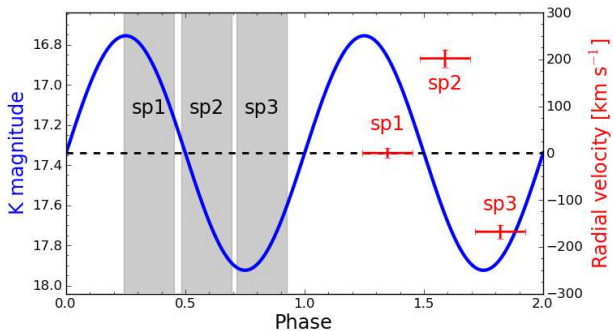}
\caption{Phased lightcurve of CBS~7. Two cycles of the $P_{2}$ period are shown. The IMACS observations are indicated by gray shaded area in the first cycle (representing the exposure time). The right axis indicates the radial velocity scale for the measurements plotted in the second cycle. Radial velocities are relative to sp1.}
\label{fig:spphase}
\end{figure}

The spectra of CBS~7 are shown in Figure~\ref{fig:3sp}. The spectra show H, HeI and HeII $\lambda$4686 broad emission lines as well as evidence for flux variability between the 3 observations. 
{The spectrum sp1 also appears slightly redder than sp3 in Figure~\ref{fig:3sp}.}
We show close-ups of the H$\beta$ and H$\alpha$ regions of the spectra after scaling the continuum to 1 in Figure~\ref{fig:3spzoom}. 
In sp3, we note the presence of emission at the wavelength of the Bowen blend (CIII/NIII $\lambda$4640).

For each spectrum and each major emission line, we report in Table ~\ref{tbl:el} the results of line fitting with a lorentzian using IRAF splot and the deblend command. The values are consistent from one line to another, except the H$\alpha$ measurement for sp2 which present a clear asymmetry in Figure~\ref{fig:3spzoom} and is therefore less reliable. The HeII $\lambda$4868 and H$\alpha$ lines appear to broaden in sp3. HeI lines measurement are affected by low signal to noise.

Given the uncertainties on the X-ray periods, the phase of our observations is indeterminate.
{If we assume that the optical flux and the NIR flux vary in phase at the period $P_{2}$}, we can fit the median optical flux of the spectra to the phased lightcurve obtained in the NIR in order to determine the possible phase of the IMACS observations. The result is reported as shaded regions in Figure~\ref{fig:spphase} where the width and the relative time of each region correspond to the exposure time and the start time of each spectrum acquisition. The variation in flux of the spectra is consistent with the flux variation in the $K_s$-band, and the error on the phase is then $\pm$0.05.
{However, it is possible that both flux variations arise from different processes as observations were performed in different bands, and at different epochs.}

We cross-correlated the 3 spectra with the first spectrum as a reference using the task fxcor in IRAF. We obtained radial velocity measurements relative to the first spectrum which are reported in Figure~\ref{fig:spphase}. 
{Though those measurements look consistent with the $P_{2}$ period, having only three points is limiting in the determination of an orbital period. If the phase of the spectroscopic observation is correct, there seems to be a phase shift of $\sim$0.25 between the flux curve and the radial velocity curve.}


\subsection{Spectroscopic identifications}\nopagebreak

\nopagebreak
We fitted all 46 spectra with a \citet{Kurucz:1993p5878} stellar atmosphere model and the \citet{Savage:1979p5573} extinction curve using specview\footnote{http://www.stsci.edu/resources/software\_hardware/specview/}. The surface gravity and metallicity of the star could not be constrained and were fixed to 4.5 (main sequence star) and the solar metallicity, respectively.

We first found that our selection of sources with an H$\alpha$ excess and no X-ray source detected (31 targets) are identified as late type stars (28 M stars, 3 K stars) with no presence of an H$\alpha$ emission line in their spectrum, except for one M star (J2000 RA=15:42:04.74 Dec=-52:27:27.9, with H$\alpha$ EW=$-5.0$ \AA, temperature 3700~K, $A_{V}$=1.1). 
We list in Table~\ref{tab:allx} the 15 targets that are aligned with an X-ray source and give X-ray and optical properties. 
We note that except for source 1 (CBS~7), the recorded net counts are low, and sometimes not sufficient for reliable flux estimates. 
In Table~\ref{tab:allsp}, we report the results of the quantile analysis for the X-ray sources (flux and absorption, \citealt{Hong:2004p5733}) and the spectral fitting of the optical spectrum (surface temperature and extinction) for the 15 targets with an X-ray counterpart (similarly to \citealt{Koenig:2008p5597}).
In this table, the extinction $A_{V}^{X}$ derived from the X-ray absorption estimate (with large errors due to the low number of counts) is consistent with the extinction $A_{V}$ derived from the optical spectrum.
{Sources 4, 5 and 8 have only 3-4 net counts, and therefore we could not derive an estimate of the absorption for those sources. Assuming a power law spectrum with a mean photon index 1.7, and the absorption derived from the extinction of their spectrum ($A_{V}$ in Table~\ref{tab:allsp}), those sources have an unabsorbed 0.5--8 keV flux of $\sim$$2\times10^{-15}$~\ergcm, and their log($F_{X}/F_{V}$)$_{u}$ is lower than --2.} 
For CBS~7, the values come from the X-ray spectral fitting (see van den Berg et al 2011) and are thus more reliable than for the other sources.

CBS~7 stands out with significant emission lines in its spectrum and a high X-ray to optical flux ratio as expected for accreting binaries \citep[e.g.][]{Verbunt:1997p1047}. 
All other targets are late type stars and show an X-ray to optical flux ratio between logarithmic values $-2$ and $-5$, consistent with stellar coronal emission \citep[e.g.][]{Schmitt:1995p5591}. 
We then note that all targets with an H$\alpha$ excess (sources 3, 6, 8, 11, 12 and 13, with negative H$\alpha-R$) and sources 10 and 15 (with H$\alpha-R=0.1\pm0.1$), show narrow Balmer emission lines (Me and Ke stars) indicating that they are chromospherically active stars with high coronal temperature. These stars are probably the optical counterparts of their X-ray source.
Other targets may be chance alignments with an X-ray source. Also, sources 5, 7 and 9 have $A_{V}^{X}$ or median energy (E50) values consistent with large absorption which may normally prevent the detection of their optical counterparts. 



\begin{deluxetable*}{ccccccccccccc}
\tablecaption{Targets aligned with an X-ray source selected for optical spectroscopy.\label{tab:allx}}
\tablewidth{0pt}
\tabletypesize{\scriptsize}
\tablehead{\colhead{ID} & \colhead{CXOPS J} & \colhead{Net Cts} & \colhead{E50} & \colhead{Err$_{95\%}$} & \colhead{Dist} &  \colhead{R.A.} & \colhead{Dec} & \colhead{$V$} & \colhead{$R$} & \colhead{$H\alpha$} & \colhead{$I$}}
\startdata
1    & 154305.5-522709 &  667.2$\pm$26.9                    & 2.2$\pm$0.1        & 0.53\arcsec & 0.34\arcsec & 15:43:05.51 & -52:27:09.6 &  21.5$\pm$0.1      &  20.9$\pm$0.2      &  20.3$\pm$0.1      &  20.3$\pm$0.1      \\
2    & 154302.5-522036 &  15.0$\pm$5.2                     & 3.0$\pm$0.5          & 2.22\arcsec & 2.81\arcsec & 15:43:02.33 & -52:20:39.3 &  \nodata    &  \nodata         &  22.7$\pm$0.3      &  \nodata          \\
3    & 154258.0-522517 &  15.6$\pm$5.3                     & 1.1$\pm$0.2          & 1.42\arcsec & 0.52\arcsec & 15:42:58.02 & -52:25:18.0 &  19.0$\pm$0.1      &  17.4$\pm$0.1      &  17.1$\pm$0.1      &  15.8$\pm$0.1      \\
4    & 154254.5-522558 &  3.8$\pm$3.6                     & 1.5$\pm$0.5           & 3.13\arcsec & 2.59\arcsec & 15:42:54.41 & -52:25:56.6 &  23.8$\pm$0.3      &  \nodata                    &  22.3$\pm$0.2      &  20.9$\pm$0.2      \\
5    & 154253.8-522017 &  3.2$\pm$3.4                     & 6.1$\pm$1.2           & 2.81\arcsec & 2.28\arcsec & 15:42:53.69 & -52:20:15.9 &  23.5$\pm$0.2      &  21.5$\pm$0.2      &  22.0$\pm$0.2      &  20.3$\pm$0.1      \\
6    & 154250.4-522152 &  36.3$\pm$7.1                     & 1.3$\pm$0.1         &  0.57\arcsec & 0.41\arcsec  & 15:42:50.35 & -52:21:52.9 &  18.6$\pm$0.1      &  17.2$\pm$0.1      &  16.9$\pm$0.1      &  15.9$\pm$0.1      \\
7    & 154240.4-522158 &  8.1$\pm$4.3                     & 2.2$\pm$0.7           &  0.66\arcsec & 0.16\arcsec  & 15:42:40.33 & -52:21:59.2 &  20.0$\pm$0.1      &  17.8$\pm$0.1      &  17.9$\pm$0.1      &  16.3$\pm$0.1      \\
8    & 154236.3-522416 &  4.2$\pm$3.4                     & 1.1$\pm$0.4           &  1.35\arcsec & 0.89\arcsec  & 15:42:36.31 & -52:24:16.5 &  21.2$\pm$0.1      &  19.6$\pm$0.1      &  19.4$\pm$0.1      &  18.2$\pm$0.1      \\
9    & 154207.6-522500 &  20.3$\pm$5.8                     & 1.9$\pm$0.2         &  0.66\arcsec & 0.47\arcsec  & 15:42:07.60 & -52:25:01.8 &  19.4$\pm$0.1      &  17.5$\pm$0.1      &  17.5$\pm$0.1      &  16.5$\pm$0.2      \\
10   & 154202.5-522701 &  10.3$\pm$4.6                     & 1.5$\pm$0.4         &  1.94\arcsec & 0.09\arcsec  & 15:42:02.43 & -52:27:02.6 &  21.8$\pm$0.1      &  19.5$\pm$0.1      &  19.7$\pm$0.1      &  17.9$\pm$0.1      \\
11   & 154201.6-521943 &  12.8$\pm$4.9                     & 1.0$\pm$0.1         &  1.17\arcsec & 0.96\arcsec  & 15:42:01.61 & -52:19:44.1 &  18.8$\pm$0.1      &  17.5$\pm$0.1      &  17.2$\pm$0.1      &  16.2$\pm$0.1      \\
12   & 154157.4-522318 &  20.3$\pm$5.8                     & 1.3$\pm$0.1         &  0.75\arcsec & 0.28\arcsec  & 15:41:57.32 & -52:23:19.3 &  19.0$\pm$0.1      &  17.9$\pm$0.1      &  17.8$\pm$0.1      &  17.0$\pm$0.1      \\
13   & 154153.1-522106 &  8.2$\pm$4.3                     & 1.2$\pm$0.3           &  1.73\arcsec & 0.96\arcsec  & 15:41:53.10 & -52:21:08.0 &  21.4$\pm$0.1      &  20.1$\pm$0.1      &  19.9$\pm$0.1      &  18.7$\pm$0.1     \\
14   & 154151.1-522658 &  8.0$\pm$4.3                     & 1.9$\pm$0.6           &  2.73\arcsec & 0.79\arcsec  & 15:41:51.16 & -52:26:59.5 &  22.2$\pm$0.1      &  20.0$\pm$0.1      &  20.0$\pm$0.1      &  18.5$\pm$0.1      \\
15   & 154150.2-522505 &  13.3$\pm$5.0                     & 2.6$\pm$0.6         &  1.90\arcsec & 0.80\arcsec  & 15:41:50.18 & -52:25:05.9 &  21.2$\pm$0.1      &  19.2$\pm$0.1      &  19.3$\pm$0.1      &  17.7$\pm$0.1      
\enddata
\tablecomments{The net count error was derived using Gehrels statistics \citep{Hong:2005p6560}. As the expected background is low, even the sources 3--4 counts are significantly detected, however the count estimate and derived flux cannot be constrained. E50 is the median energy in keV of the X-ray photons in the 0.5--8~keV range; coordinates R.A. ($h$:$m$:$s$) and Dec ($^{\circ}$:\arcmin:\arcsec) are for the optical counterpart.}
\end{deluxetable*}

\begin{deluxetable*}{ccccccccll}
\tablecaption{Properties of targets aligned with an X-ray source from X-ray and optical data.\label{tab:allsp}}
\tablewidth{0pt}
\tabletypesize{\scriptsize}
\tablehead{\colhead{ID} & \colhead{$F_{X,u}$} & \colhead{$N_{\rm H}$} & \colhead{$A_{V}^{X}$} & \colhead{$A_{V}$} & \colhead{Temp} & \colhead{log($F_{X}/F_{V}$)$_{u}$}  & \colhead{H$\alpha-R$} & \colhead{Type} & \colhead{Notes}}
\startdata
1    &  51.0             $\pm$ 2.0                     &  0.24         $\pm$ 0.02    &  1.3   $\pm$ 0.2      & \nodata &  \nodata  & \textbf{1.4\,$\pm$\,0.2}  &  \textbf{$-$0.5}    $\pm$ 0.2        & CV   &  CBS~7, emission lines (see Table~\ref{tbl:el})       \\
2    &  1.5               $\pm$ 0.5                     &  0.52         $\pm$ 0.66   &  2.9   $\pm$ 3.7      & \nodata    & \nodata     & \nodata &  \nodata  &      &    no detection in spectroscopic data         \\
3    &  1.2               $\pm$ 0.4                     &  0.29         $\pm$ 0.15   &  1.6   $\pm$ 0.8      & 0.9 & 3300 & $-4.5\pm0.4$ &  \textbf{$-$0.3}    $\pm$ 0.1        & M5e  & H$\alpha$ EW=$-8.3$ \AA               \\
4    &  \nodata                    &  \nodata   &  \nodata     & 1.9 & 5000 & \nodata &  \nodata          & K2   &                         \\
5    &  \nodata                 &  \nodata &  \nodata    & 1.5  & 4800 & \nodata  &  $0.5$     $\pm$ 0.3        & K4   & spectrum truncated above 6500 \AA             \\
6    &  1.3               $\pm$ 0.3                     &  0.03         $\pm$ 0.06   &  0.2   $\pm$ 0.3      & 0.6  & 3500 & $-4.5\pm0.3$ &  \textbf{$-$0.3}    $\pm$ 0.1        & M3e  & H$\alpha$ EW=$-5.4$ \AA               \\
7    &  0.8               $\pm$ 0.4                     &  0.98         $\pm$ 0.94   &  5.4   $\pm$ 5.2      & 2.6 & 4500 & $-5.5\pm0.6$ &  0.1     $\pm$ 0.1        & K5   &                         \\
8    &  \nodata                  &  \nodata   &  \nodata      & 1.9  & 4100 & \nodata &  $-$0.2    $\pm$ 0.1        & M0e  & H$\alpha$ EW=$-2.0$ \AA \\
9    &  14.5              $\pm$ 4.1                     &  2.2          $\pm$ 1.0   &  12.3  $\pm$ 5.5      & 1.6 & 4700 & $-2.3\pm0.4$ &  0.0     $\pm$ 0.1        & K4   &                         \\
10   &  0.5               $\pm$ 0.2                     &  0.09         $\pm$ 0.16   &  0.5   $\pm$ 0.9      & 2.0 & 4500 & $-3.8\pm0.5$ &  0.1     $\pm$ 0.1        & K5e  & H$\alpha$ EW=$-1.2$ \AA  \\
11   &  7.8               $\pm$ 2.9                     &  0.78         $\pm$ 0.30    &  4.3   $\pm$ 1.7      & 1.2 & 3500 & $-3.0\pm0.4$ &  \textbf{$-$0.3}    $\pm$ 0.1        & M3e  & H$\beta$ EW=$-0.8$ \AA / truncated 6000 \AA        \\
12   &  1.1               $\pm$ 0.3                     &  0.22         $\pm$ 0.55   &  1.2   $\pm$ 3.1      & 0.6 & 4600 & $-4.3\pm0.4$ &  $-$0.1    $\pm$ 0.1        & K4e  & H$\alpha$ EW=$-0.8$ \AA \\
13   &  0.3               $\pm$ 0.2                     &  0.01         $\pm$ 0.08   &  0.1   $\pm$ 0.5      & 0.5 & 3900 & $-3.2\pm0.6$  &  $-$0.2    $\pm$ 0.1        & M1e  & H$\alpha$ EW=$-2.9$ \AA               \\
14   &  0.6               $\pm$ 0.3                     &  0.35         $\pm$ 0.61   &  1.9   $\pm$ 3.4      & 1.8 & 4300 & $-3.1\pm0.6$ &  0.1     $\pm$ 0.1        & K6   &  \\
15   &  2.2               $\pm$ 0.8                     &  1.75         $\pm$ 1.36   &  9.8   $\pm$ 7.6      & 1.9 & 4500 & $-2.8\pm0.4$ &  0.1     $\pm$ 0.1        & K5e   & H$\alpha$ EW=$-1.5$ \AA
\enddata
\tablecomments{$F_{X,u}$ is the unabsorbed 0.5--8~keV flux in $10^{-14}$~\ergcm\ -- {sources 4, 5 and 8 have only 3-4 counts and thus no reliable flux estimates}; $N_{\rm H}$ is the absorption in $10^{22}$~cm$^{-2}$; $A_{V}^{X}$ is the extinction derived from the absorption with $A_{V}^{X}=N_{\rm H} / 1.79\times 10^{21}$ \citep{Predehl:1995p5533}; $A_{V}$ and the star temperature in K (Temp) are obtained from the optical spectrum fit using Kurucz stellar models, with typical errors of $\pm$0.2 and $\pm$200, respectively; log($F_{X}/F_{V}$)$_{u}$ is the X-ray to optical V-band unabsorbed flux ratio, and Type gives the identified spectral type based on the temperature or features in the spectrum (equivalent width, EW, are given for emission lines). H$\alpha-R$ values in bold indicate the selection of 4 targets with H$\alpha$ excess from photometry.} 
\end{deluxetable*}


\section{Discussion}
\label{discuss}

Using multi-wavelength observations, we identify 8 X-ray sources as late-type stars with emission lines. This is consistent with the correlation observed for late-type stars between strong chromosphere activity \citep{Stauffer:1986p5590} which produces Balmer narrow emission lines, and the presence of a hot corona which emits X-rays \citep{Fleming:1995p5586,Schmitt:1995p5591,Gudel:2004p5596}. Another 4 sources without emission lines could be chance alignments, and one last is a probable AGN. Those results are consistent with the results of \citet{Koenig:2008p5597}, where most of the identified X-ray sources appear to be single stars.
Based on optical spectra and NIR variability and colors, we can further discuss the nature of a candidate CV, CBS~7, that showed a hard X-ray spectrum, a periodic lightcurve, and a H$\alpha$ excess of its optical counterpart (van den Berg et al. 2011). 

We clearly show here that CBS~7 is a CV through the detection of H, HeI and HeII emission lines which are consistent with the presence of accretion as observed for many CVs \citep[e.g.][]{Williams:1983p4480}. 
{The Bowen blend emission lines have been noticed in accreting binary systems and could be related to the irradiated companion star \cite[e.g.][]{Steeghs:2002p7442}.}
The HeII $\lambda$4686 line in emission indicates that there is a cloud of material exposed to a strong enough ionizing continuum shortward of the HeII $\lambda$228\AA\ edge. Such an emission could come from a region close to the magnetic pole of a WD, indicating a MCV.
Following the empirical test of \cite{Silber:1992p4495} --- given that the EW of H$\beta$ is higher than 20 \AA\ and $F_{HeII}/F_{H\beta}$ is greater than 0.4 --- CBS~7 could indeed be an IP.
The optical spectrum is also consistent with the spectra of known polars \citep[e.g][]{Schwarz:2002p4737}.
{Alternatively, a HeII $\lambda$4686 line in emission is common in SW Sex nova-like CVs and post-nova CVs \citep{Thorstensen:1991p7431,RodriguezGil:2005p7432}.} 

{However, the hard \xmm\ X-ray spectrum of CBS~7 with a temperature $>$30~keV is not consistent with the system being a nova-like, and more resembles the X-ray emission of MCVs.
Indeed, non-magnetic CVs generally have softer X-ray emission (temperature of a few keV, \citealt{Baskill:2005p960}), while MCVs can be harder \citep[tens of keV,][]{Heinke:2008p4733}. The luminosity of CBS~7 corresponds to the highest luminosity of non-magnetic CVs, and matches well the expected luminosity of MCVs \citep{Verbunt:1997p1047}. We note that there could be internal absorption in CVs which can cause our distance and luminosity estimates to be overestimated.}

{The X-ray and near-infrared variability indicate that the period $2.39\pm0.05$~h could be a fundamental period of the system. 
The three radial velocity measurements are compatible with such a period, though they are not very constraining, given that we assumed that NIR and optical emissions were varying in phase. Also, the non-simultaneous observations could have caught the source in a different state.
Nevertheless, it is possible that $P_{2}$ is the orbital period of the system.}
The system would then fall in the CV period gap, between 2 and 3~h \citep{Spruit:1983p4497}. 
We note that the period gap is less significant for MCVs, which would bring further support to the idea that CBS~7 is a MCV. 

{From the possible orbital period of CBS~7, the expected type of the companion is around dM4$\pm$1 following the standard evolution model described e.g. by \citet{Knigge:2011p4724}. 
Based on the $JHK_s$ colors (Figure~\ref{fig:ccd}), the nature of the companion star is less constrained as they are compatible with a $\sim$dM3 star or less massive star \citep[see][for comparisons with other CVs]{Hoard:2002p4706}.}
The predicted absolute magnitudes of donor stars in the NIR can be used to set a lower limit (as the accretion disk also contributes to the emission) on the distance of the system \citep[e.g.][]{Knigge:2011p4724}. Using $M_{K_s}$=8 for a dM5 star and $K_s$=18 (lower value observed), we obtain $d \gtrsim 1.0$~kpc.
{If the companion star is a less massive star, then this limit can be lower, leading us to give a rough estimate of the distance of $\sim$1.0~kpc}, which is consistent with the estimate of the distance reported by van den Berg (2011) of $\sim$1.4~kpc.


IPs are generally identified through the presence of two periods in their X-ray lightcurve. The fact that two periods are detected for CBS~7 could therefore suggest an IP nature. The ratio $P_{spin}/P_{orb}$ would then be $\sim$0.5. 
Such a ratio is high for an IP \citep[see e.g.][]{Norton:2008p4760} and could indicate a near-synchronous IP, a rare class of MCVs.
Compared to IP and polar populations (e.g. from \citealt{Heinke:2008p4733} and \citealt{Norton:2008p4760}), CBS~7 has a similar X-ray luminosity to the brightest polars and the faintest IPs, and an orbital period compatible with both populations. 
CBS~7 could therefore possibly be a transition object between IPs and polars, and a new member of the rare population of near-synchronous IPs (6 objects known).
According to the evolutionary model of MCVs \citep{Norton:2008p4760}, IPs start with $P_{spin}/P_{orb} < 0.1$ and as the systems evolve through magnetic lock, the orbital periods decrease and the spin periods increase, i.e. $P_{spin}/P_{orb}$ approaches 1. Therefore, the orbital periods of near-synchronous IPs are clustered around or below the period gap near the end of the evolution, where CBS~7 is located.
This population may be more numerous on a Galactic scale than previously recognized, as we recently identified 10 periodic MCVs close to the Galactic Center \citep{Hong:2011p4354} with similar properties (hard X-ray spectrum, periods at or below the period gap).

Alternatively, a ratio of 0.5 for the two periods could be explained if one period is the harmonic of the other. As no other period appeared to be significant down to 20~s (van den Berg et al. 2011), CBS~7 does not seem to be an asynchronous IP with a period ratio $<$0.1. However, it could be a polar with locked spin and orbital periods. Secondary X-ray dips have been observed in some polars and could come from the occultation of the emitting pole either by the WD itself or by the accretion column \citep{Mason:1985p4761}.
If the system is accreting from 2 poles, then we might see alternatively one of these poles while the other is smoothly occulted.
The phased lightcurve of polars can be variable over time \citep{Osborne:1987p4765}, and in some cases is similar to the observed lightcurve of CBS~7. 
A search for polarized optical emission, as well as a better sampling of the phased lightcurve in X-rays to determine other periods, could confirm this object as either an IP, a near-synchronous IP or a polar.


\section{Conclusions}

We presented multi-wavelength follow-up observations of an archived \chandra\ observation from the ChaMPlane survey. Among the X-ray sources, one is identified as a CV (CBS~7), 8 as active late-type stars, 5 may be chance alignments. The stars with an H$\alpha$ excess from photometry not corresponding to Chandra sources are all identified as late type stars.

{The source CBS~7, selected from its X-ray emission and an H$\alpha$ excess is clearly identified as a CV through optical spectroscopy. It is possibly an IP, a near-synchronous IP or a polar. Both properties (X-ray emission and H$\alpha$ excess) thus appear to be essential in the selection of accreting compact binary candidates, but are not sufficient to clearly identify them (one CV among 4 candidates in this work). The detection of variability as a third property seems more conclusive (see also \citealt{Hong:2009p2775}).
The use of those criteria in the ChaMPlane survey thus led to a majority of targets identified as foreground, late-type, active stars. However, and more importantly, we identify a few rare accreting compact binaries at intermediate distance in the Galactic Plane such as CBS~7 ($d \gtrsim 1.0$~kpc), which are crucial in order to better understand the Galactic accreting compact binary population.}


\acknowledgments

We thank the anonymous referee for useful comments that helped improve the paper. 
We acknowledge support from NASA/\chandra\ grants AR9-0013X, GO9-0102X and NSF grant AST-0909073. We thank the Chandra X-ray Center (CXC) for support. 
STSDAS, PyRAF and Specview are products of the Space Telescope Science Institute, which is operated by AURA for NASA.



{\it Facilities:} \facility{Magellan}, \facility{CXO (ACIS)}.






\bibliographystyle{apj} 
\bibliography{../../../ref.bib}

\end{document}